\documentclass[preprint,aps]{revtex4}
\usepackage{mathrsfs}
\usepackage{graphicx}
\usepackage{multirow}
\usepackage{makecell}
\usepackage{booktabs}
\usepackage{bm}
\usepackage{bbm}
\begin{document}

\title{Spectroscopic Evidence for Dirac Nodal Surfaces and Nodal Rings in  Superconductor NaAlSi}

\author{Chunyao Song$^{1,2\sharp}$, Lei Jin$^{3,4\sharp}$, Pengbo Song$^{1,2\sharp}$, Hongtao Rong$^{1,2}$, Wenpei Zhu$^{1,2}$, Bo Liang$^{1,2}$, Shengtao Cui$^{5}$, Zhe Sun$^{5}$, Lin Zhao$^{1,2,7}$, Youguo Shi$^{1,2,7}$, Xiaoming Zhang$^{3,4*}$, Guodong Liu$^{1,2,7*}$ and X. J. Zhou$^{1,2,6,7*}$ }

\affiliation{
	\\$^{1}$Beijing National Laboratory for Condensed Matter Physics, Institute of Physics, Chinese Academy of Sciences, Beijing 100190, China
	\\$^{2}$University of Chinese Academy of Sciences, Beijing 100049, China
	\\$^{3}$State Key Laboratory of Reliability and Intelligence of Electrical Equipment, Hebei University of Technology, Tianjin 300130, China
	\\$^{4}$School of Materials Science and Engineering, Hebei University of Technology, Tianjin 300130, China
	\\$^{5}$National Synchrotron Radiation Laboratory, University of Science and Technology of China, Hefei 230029, China
	\\$^{6}$Beijing Academy of Quantum Information Sciences, Beijing 100193, China
	\\$^{7}$Songshan Lake Materials Laboratory, Dongguan 523808, China
	\\$^{\sharp}$These people contributed equally to the present work.
	\\$^{*}$Corresponding authors:XJZhou@iphy.ac.cn, gdliu$\_$ARPES@iphy.ac.cn and zhangxiaoming87@hebut.edu.cn
}

\date{\today}


\begin{abstract}
The discovery of the topological states has become a key topic in condensed matter physics with the focus evolving from the Dirac or Weyl points to high-dimension topological states of the nodal lines and nodal surfaces. For a topological material to manifest its quantum properties and become useful in applications, the topological states need to be genuine and clean so that they lie close to the Fermi level without other trivial bands existing at the Fermi level. While a number of high-dimension topological materials are predicted, only a few of them have been synthesized and confirmed and the genuine and clean ones are especially scarce. Here we report the realization of the genuine clean multiple high-dimension topological states in NaAlSi. By performing high-resolution angle-resolved photoemission measurements and band structure calculations, we have observed two sets of nodal surfaces and the formation of two homocentric nodal ring states in NaAlSi. The observed nodal rings are distinct in that the inner one is a type-{\uppercase\expandafter{\romannumeral1}} nodal ring while the outer one is a type-{\uppercase\expandafter{\romannumeral1}} nodal ring embedded with four type-{\uppercase\expandafter{\romannumeral3}} nodal points. All the bands involved in the nodal rings lie very close to the Fermi level with no other trivial bands coexisting at the Fermi level. These observations make NaAlSi a desirable topological material to explore for novel quantum states and exotic properties.

\end{abstract}

\maketitle
Searching for topological materials has become one of the central topics in condensed matter physics because they exhibit novel topological states, exotic properties and a great potential of applications\cite{Hasan_RevModPhys_2010_Topologicalinsulator,Qixiaoliang_RevModPhys_2011_Topologicalinsulatorsandsuperconductors,Armitage_RevModPhys_2018_WeylandDiracsemimetalsinthree-dimensionalsolids,SatoMasatoshi_ReportsonProgressinPhysics_2017_Topologicalsuperconductorsareview,TokuraYoshinori_NatRevPhy_2019_Magnetictopologicalinsulators}. The studied materials have evolved from topological insulators to Dirac/Weyl semimetals to topological superconductors and to topological magnets, the topological states have developed from Dirac states to Weyl states to Majorana states and to high-order topological states. The dimension of the topological states also progresses from zero-dimensional points to one-dimensional lines to two-dimensional surfaces because higher dimension topological states are expected to show unique properties\cite{HuhYejin_PRB_2016_Coulombinteraction,MitchellAndrewK_PRB_2015_Kondoeffect,RhimJun-Won_PRB_2015_Landaulevelquantization,RamamurthySrinidhiT_PRB_2017_Quasitopologicalelectromagneticresponse}. In practice, to fully achieve the topological properties, the topological states need to be genuine so that they are robust against spin-orbital coupling and lie close to the Fermi level\cite{Wangyang_PRB_2021_LaSbTe}. The topological materials also need to be clean so that the electronic states near the Fermi level are dominated by topological states. While a number of high-dimension topological states have been predicted\cite{Chen_ACS_2015_Carbonallotropes_Prediction,Kieran_PRL_2015_trigonallyconnected3Dlattices_Prediction,Weng_PRB_2015_graphenenetworks_Prediction,Zeng_arXiv_2015_LaX_Prediction,Kim_PRL_2015_metal-dopedCu3N_Prediction,Yu_PRL_2015_Cu3PdN_Prediction,Xu_PRB_2015_ZrSiO_Prediction,Xie_PRB_2015_Ca3P2_Prediction,Ai_PSJ_2015_CaAgAs_Prediction,Wang_PRL_2016_C16_Prediction,Li_PRL_2016_alkaliearthmetals_Prediction,Zhu_PRB_2016_Ba2X_Prediction,Huang_PRB_2016_BaSn2_Prediction,Liang_PRB_2016_BaMX3_Prediction,Zhao_PRB_2016_blackphosphorus_Prediction,Bian_NC_2016_PbTaSe2,Du_npj_2017_CaTe_Prediction,Xu_PRB_2017_CaP3_Prediction,Takahashi_PRB_2017_Al3FeSi2_Prediction,Hirayama_NC_2017_alkalineearthmetals_Prediction,Li_SCM_2018_TiSi_Prediction,Facio_PRM_2019_Pt2HgSe3_Prediction,Ghosh_PRB_2019_Pt2HgSe3_Prediction}, only a few of the nodal lines and nodal surfaces have been realized\cite{Schoop_NC_2016_ZrSiS,Chen_PRB_2017_HfSiS,Fu_SciAdv_2018_ZrSiS,Bian_NC_2016_PbTaSe2,Takane_npj_2018_CaAgAs,Song_PRL_2020_SrAs3,Takafumi_PRB_2018_Ta3SiTe6,Wu_NP_2016_PtSn4,Cucchi_PRL_2020_Pt2HgSe3,Ekahana_NJP_2017_InBi,Feng_NC_2017_monolayerCu2Si,Feng_PRL_2019_monolayerGdAg2} and the genuine and clean high-dimension topological states have been particularly scarce.  Moreover, the topological superconductivity has recently attracted considerable attention due to the feasibility of realizing topological qubits for fault-tolerant quantum computation\cite{SatoMasatoshi_ReportsonProgressinPhysics_2017_Topologicalsuperconductorsareview}. However, up to date, the discovered topological superconductors are very rare. The coexistence of intrinsic superconductivity and topological phases in one and the same material can give us opportunities to find new and more practical topological superconductors and thus realize Majorana zero modes. 

In this paper, we report discovery of multiple topological states in NaAlSi which is recently predicted to be a topological material candidate\cite{LeiJin_JMC_2019_NaAlSiCal,Yi_JMCC_2019_NaAlSiCal,Muechler_APL_2019_NaAlSiCal}. NaAlSi is quite intriguing semimetal that is composed of light sp electron elements, intrinsic self-doped with low carrier density, and superconducting at $\sim$7K possibly with unconventional pairing mechanism\cite{Kuroiwa_PhysicaC_2007_Superconducting,Yamada_JPSJ_2021_Superconducting}. It crystallizes in the same  structure as the iron-pnictide ``111'' compounds. Therefore, combined with the intrinsic bulk superconductivity and topological nodal ring states, NaAlSi might provide a fascinating platform to investigate topological superconductivity. By carrying out high resolution angle resolved photoemission (ARPES) measurements, we have observed topological nodal surfaces and nodal rings states in NaAlSi. The electronic states near the Fermi level are dominated by the bands that form the topological states and the formed topological rings lie close to the Fermi level. These results indicate that NaAlSi is a clean and genuine topological material hosting novel topological states and exotic physical properties.

High-quality single crystals of NaAlSi were grown by flux method, as described in Supplemental Materials\cite{SM}. ARPES measurements were performed at BL-13U in National Synchrotron Radiation Laboratory (NSRL, Hefei, China), with a Scienta Omicron DA30L analyzer. The total energy and angular resolution were $\sim$7\,meV and $\sim$0.3 degree, respectively. In order to get complete electronic structure, two polarization geometries were used. In the LH polarization geometry, the vector of the electric field of the incident light is parallel to the horizontal plane, while in the LV polarization geometry, it is perpendicular to the horizontal plane. NaAlSi single crystal samples were cleaved {\it in situ} and measured in ultrahigh vacuum better than 8.0$\times$10$^{-11}$ mbar. Our first-principles band-structure calculations based on the framework of density functional theory (DFT) have been done using the Vienna ab {\it initio} simulation package (VASP) within the generalized gradient approximation (GGA) of Perdew-Burke-Ernzerhof type and Heyd–Scuseria–Ernzerhof (HSE06) type\cite{HES06_Calculations_2003}. The cutoff energy is set as 400eV. A $\Gamma$-centered k-mesh of 13 $\times$ 13 $\times$ 7 is used for the Brillouin zone sampling. The optimized lattice parameters a=b=4.139 $\AA$, c=7.310 $\AA$ are employed.

NaAlSi crystallizes in a non-symmorphic space group P4/nmm (NO.129)\cite{Westerhaus_ZNB_1979_NaAlSicrystal}; its crystal structure is illustrated in Fig. 1a which is isostructural to that of the ZrSiS and LaSbTe family\cite{Schoop_NC_2016_ZrSiS,Xu_PRB_2015_ZrSiO_Prediction,Wangyang_PRB_2021_LaSbTe}. Fig. 1c shows the calculated band structures of NaAlSi with and without considering the spin-orbital couplings (SOC). The electronic states near the Fermi level are dominated by three bands, labeled as $\alpha$, $\beta$ and $\gamma$ in Fig. 1c which are mainly contributed by the Si-3p and Al-3s orbitals\cite{LeiJin_JMC_2019_NaAlSiCal,Yi_JMCC_2019_NaAlSiCal,Muechler_APL_2019_NaAlSiCal}. The light elements in NaAlSi are expected to give rise to weak SOC effect (less than 10\,meV)\cite{LeiJin_JMC_2019_NaAlSiCal}, as evidenced by the small difference between the bands with and without SOC in Fig. 1c. These three bands near the Fermi level also exhibit a quasi-two-dimensionality (2D) as demonstrated by the weak band dispersion along the $\Gamma$-Z direction in Fig. 1c. 

Figure 1d-1h shows the band structures measured along two high-symmetry directions with different photon energies and under different polarization geometries. The band structures measured with many different photon energies are shown in Fig. S2 in Supplemental Material\cite{SM}. Different k$_{z}$s are measured by using different photon energies with the photon energies of 36\,eV and 26\,eV corresponding to k$_{z}$  $\sim0$ and k$_{z}$$\sim\pi$/c, respectively. From the measurement along $\bar\Gamma$-$\bar{X}$ with 26\,eV photon energy under the LH polarization in Fig. 1d, three bands can be clearly observed and labeled as $\alpha$, $\beta$ and $\gamma$. Here the top part of the $\beta$ band and the bottom part of the $\gamma$ band are suppressed due to the photoemission matrix element effect. When the polarization is switched to LV in Fig. 1e, the full $\beta$ band can be observed while the other $\alpha$ and $\gamma$ bands are strongly suppressed. Therefore, measurements under different polarization geometries provide a more complete band structure. Fig. 1f shows the band structure measured along $\bar\Gamma$-$\bar{X}$ with a photon energy of 36\,eV (k$_{z}$$\sim$0). Compared with the measurements with the 26\,eV photon energy (k$_{z}$$\sim\pi$/c) in Fig. 1d and 1e, the observed $\alpha$ and $\beta$ bands shift upwards by 0.14\,eV while the $\gamma$ band moves downwards by 0.05\,eV. Fig. 1g and 1h shows the band structures measured along $\bar\Gamma$-$\bar{M}$ under two different polarization geometries. Compared with the band structures measured along $\bar\Gamma$-$\bar{X}$ in Fig. 1d and 1e, the observed $\alpha$ and $\beta$ bands come close to each other while they are far apart in the $\bar\Gamma$-$\bar{X}$ measurements.

The measured band structures of NaAlSi are in good agreement with the band structure calculations. In Fig. 1d-1h, we also plot the calculated band structures of NaAlSi on top of the measured results. By shifting the calculated bands 0.1\,eV upwards, the calculated bands can well match the measured ones without invoking any further band renormalization. Such a relative band shift may be caused by carrier doping from nonstoichiometry or defects in the measured sample. The consistency between the measurements and calculations is reflected in several aspects. First, the number, shape and positions of the measured and calculated bands, and their evolution from $\bar\Gamma$-$\bar{X}$ (Fig. 1d-1e) to $\bar\Gamma$-$\bar{M}$ (Fig. 1g-1h), agree well with each other. Second, the band dispersion along k$_{z}$ direction is clearly observed and well described. As seen in Fig. 1d and 1f, when k$_{z}$ changes from $\pi$/c (for 26\,eV photon energy in Fig. 1d) to 0 (for 36\,eV photon energy in Fig. 1f), the upward shift of the $\alpha$ and $\beta$ bands and the downward shift of the $\gamma$ band are perfectly captured by the band structure calculations. The detailed comparison between measured and calculated band structures gives us a fairly accurate description of the electronic states and an opportunity to confirm each other in NaAlSi.

The electronic structures of NaAlSi near the Fermi level are dominated by three bands: the hole-like $\alpha$ and $\beta$ bands and electron-like $\gamma$ band (Fig. 1d-1h). Although only the $\gamma$ band crosses the Fermi level to form the Fermi surface, the $\alpha$ and $\beta$ bands lie also very close to the Fermi level, particularly for the k$_{z}$=0 case (Fig. 1f). Moreover, the $\alpha$ band is flat over a wide momentum space along the $\bar\Gamma$-$\bar{X}$ direction (Fig. 1d-1f). Such a flat band can cause strong electron correlation and singularity of density of states close to Fermi level, which may enhance the superconducting pairing\cite{NaAlSi_Superconductinggap_2010,Yi_JMCC_2019_NaAlSiCal}. The crossing of the $\alpha$ flat band with the $\gamma$ band along $\bar\Gamma$-$\bar{X}$ forms Dirac structure that is characteristic of the type-\uppercase\expandafter{\romannumeral3} Dirac point\cite{Volovik_JETPLetters_2016_type3,Huang_PRB_2018_type3,Yi_JMCC_2019_NaAlSiCal}. The observation of the type-\uppercase\expandafter{\romannumeral3} Dirac point in superconducting NaAlSi makes it an interesting system to explore for novel phases and exotic properties\cite{Leykam_APX_2018_flatbands,	Li_PRB_2017_flatbands,Yi_JMCC_2019_NaAlSiCal}. 

Now we look at the electronic structure of NaAlSi from the the Fermi surface and constant energy contours point of view. Fig. 2a and 2b show the Fermi surface and constant energy contours at different binding energies measured at two photon energies of 26\,eV and 32\,eV that correspond to the k$_{z}$ planes of $\sim\pi$/c and $\sim0.37\pi$/c, respectively. Overall, the measured results in Fig. 2a and 2b at different k$_{z}$s are quite similar, consistent with the quasi-two-dimensionality of the involved three bands from the band structure calculations (Fig. 1c) and measurements at different photon energies (Fig. S2 in Supplemental Material\cite{SM}). The measured Fermi surface and constant energy contours in Fig. 2a and 2b agree well with the calculations as shown in Fig. 2c. The Fermi surface of NaAlSi consists of a single electron-like pocket that originates from the $\gamma$ band. It shrinks in its area with increasing binding energies. On the other hand, the constant energy contours consist mainly of two perpendicular elliptical pockets. Their areas increase with increasing binding energies, consistent with their origination from the two hole-like $\alpha$ and $\beta$ bands. We note that, due to the photoemission matrix element effect, the shape of the measured Fermi pocket around $\bar\Gamma$ looks different from the calculated Fermi pocket. As shown in Fig. 1d and Fig. 1g, the spectral weight of the $\gamma$ band is suppressed along the $\bar\Gamma$-$\bar{M}$ direction (Fig. 1g) when compared with that measured along the $\bar\Gamma$-$\bar{X}$ direction (Fig. 1d). This makes the spectral weight along the measured Fermi pockets in Fig. 2a and Fig. 2b not homogenous. But the extracted shape and size of the measured Fermi pocket is consistent with the calculations.

In Fig. 3, we present the direct observation of the Dirac nodal surfaces in NaAlSi. Based on the lattice symmetry analysis\cite{Fang_CPB_2016_nodallinereview,Yang_APX_2018_nodallinereview} and the band structure calculations, under the combined operation of non-symmetric screw symmetries $\tilde{\mathcal{C}}_{2x}=\left\{\mathcal{C}_{2x}|\frac{1}{2}00\right\}$, $\tilde{\mathcal{C}}_{2y}=\left\{\mathcal{C}_{2y}|0\frac{1}{2}0\right\}$ and time-reversal symmetry $\mathcal{T}$, when ignoring the SOC effect, there are degenerate band crossings filled on the Brillouin Zone (BZ) boundary X-R-A-M which makes X-R-A-M an intrinsic Dirac nodal surface as indicated by the yellow plane in Fig. 3e. When the SOC effect is considered, a small gap on the order of $\sim$10\,meV will be opened along the M-X and R-A directions (Fig. 3h) which changes the initial nodal surface in X-R-A-M degeneration plane into the nodal lines along the X-R and A-M directions. Band structure calculations indicate that, in the energy range within 4\,eV below the Fermi level, there are two sets of nodal surfaces as seen by the two Dirac points at the X and R points (D1 and D2 in Fig. 3f). The band dispersion of the two nodal surfaces is calculated and shown in Fig. 3h.

Figure 3b shows the band structures of NaAlSi measured along $\bar\Gamma$-$\bar{X}$ with different photon energies which correspond to different k$_{z}$s. These measurements represent different momentum cuts in the $\Gamma$-Z-R-X plane, as shown in Fig. 3e. In these band structures, an obvious feature is the Dirac linear crossings that lie at $\sim$1.3\,eV below the Fermi level at the $\bar{X}$ point, as marked by the red squares in Fig. 3b. The energy position of these Dirac points at different k$_{z}$s along the X-R line is plotted in Fig. 3h (red squares). It changes little with k$_{z}$ and agrees well with the calculated result that the band corresponding to D1 crossing points along X-R is nearly flat.

To demonstrate the existence of the nodal surface in the X-R-A-M plane, in addition to the measurements along X-R as shown in Fig. 3b, more measurements are necessary along the $\bar{X}$-$\bar{M}$ direction. Fig. 3d shows the band structures measured parallel to $\bar\Gamma$-$\bar{X}$ at different k$_{y}$s; the location of these momentum cuts lies in the $\bar\Gamma$-$\bar{X}$-$\bar{M}$ plane at k$_{z}$ $\sim0.37\pi$/c as shown in Fig. 3e. In the band structure measured along the B1 momentum cut (left panel in Fig. 3d), two Dirac points are clearly observed at the binding energies of 1.52\,eV and 2.43\,eV that correspond to the D1 and D2 in Fig. 3f and 3h. With the momentum cuts moving from $\bar{X}$ to $\bar{M}$, both Dirac points shift to higher binding energy and the D2 Dirac point moves out of the detection window. The Dirac crossings observed in Fig. 3d are located exactly at the same k$_{x}$ as $\bar{X}$, namely in the yellow X-R-A-M plane in Fig. 3e. These observations are consistent with the calculated results in Fig. 3h where the two Dirac points move downwards from $\bar{X}$ to $\bar{M}$ (X-M and R-A in Fig. 3h). Fig. 3g shows the band structures measured along the momentum cuts that are parallel to the X-R-A-M plane. For the momentum cut A1 in the X-R-A-M plane, the observed band (D1 in the left panel in Fig. 3g) is degenerate. For the momentum cut A2 that is away from the X-R-A-M plane, the original D1 band splits into two bands (D1Up and D1Dn in the right panel in Fig. 3g). Our measured results in Fig. 3b, 3d and 3g, combined with the band structure calculations in Fig. 3f and 3h, have provided strong evidence on the existence of two sets of the nodal surfaces in NaAlSi.

Now we present our results in Fig. 4 on the observation of the Dirac nodal rings in NaAlSi. Band structure calculations have predicated the formation of nodal rings on the k$_{z}$=0 and k$_{z}$=$\pi$/c planes\cite{LeiJin_JMC_2019_NaAlSiCal,Yi_JMCC_2019_NaAlSiCal}. These nodal rings are protected by two sets of symmetries, the simultaneous presence of time-reversal and inversion symmetries or mirror symmetry. Fig. 4e shows the calculated band structures for the momentum cuts on the k$_{z}$=$\pi$/c plane varying from ZA to ZR. The $\alpha$ band exhibits a strong momentum dependence, changing from a steep band along ZA (leftmost panel in Fig. 4e) to the appearance of the flat band along ZR (rightmost panel in Fig. 4e). The momentum dependence of the $\beta$ and $\gamma$ bands is relatively weak. For each momentum cut in Fig. 4e, the $\gamma$ band crosses the $\alpha$ and $\beta$ bands, forming two Dirac points. This gives rise to the formation of two Dirac nodal rings, as plotted in Fig. 4f, with the inner A1 ring derived from the crossings of  the $\beta$ and $\gamma$ bands and the outer A2 ring from the crossings of $\alpha$ and $\gamma$ bands.

Our measured results have confirmed the formation of the Dirac nodal rings in NaAlSi. Fig. 4b shows the measured band structures of NaAlSi along different momentum cuts from $\bar\Gamma$$\bar{M}$ to $\bar\Gamma$$\bar{X}$ direction. The photon energy used is 26\,eV that corresponds to a k$_{z}$=$\pi$/c. The observed band structures are highly consistent with the calculated results in Fig. 4e. This provides us confidence in determining the positions of all three bands although the $\gamma$ band is not clearly observed for some momentum cuts due to the photoemission matrix element effects. The crossings of the $\gamma$ band with the $\alpha$ and $\beta$ bands lead to the formation of the NR2 and NR1 Dirac nodal rings in Fig. 4c which resembles the calculated results in Fig. 4f. Our measured results in Fig. 4b and 4c, combined with the band structure calculations in Fig. 4e and 4f, strongly indicate the formation of Dirac nodal rings in NaAlSi.

In summary, by performing high-resolution ARPES measurements and band structure calculations, we have observed multiple topological states in NaAlSi. Two sets of the nodal surfaces have been observed. In addition, We directly uncovered the formation of the two homocentric nodal ring states  around the zone center at k$_{z}$=0 and $\pi$/c planes. The observed nodal rings are distinct in that the inner one is a type-{\uppercase\expandafter{\romannumeral1}} nodal ring while the outer one is a type-{\uppercase\expandafter{\romannumeral1}} nodal ring embedded with four type-{\uppercase\expandafter{\romannumeral3}} nodal points. Since the SOC effect is weak due to the light-elements in NaAlSi, the gap opening at the Dirac points is small, making NaAlSi close to a genuine topological system. It is also a clean topological material because all the three bands are involved in the formation of the nodal rings that lie very close to the Fermi level without other trivial bands coexisting at the Fermi level. Considering that NaAlSi is also superconducting, our observations of the genuine clean multiple high-dimension topological states make it a desirable platform to explore for novel phases and exotic properties.

\hspace*{\fill}

\vspace{3mm}

\noindent {\bf Acknowledgement}\\
We thank Youting Song for the measurement of X-ray diffraction and Dong Li, Xiaoli Dong for the magnetic measurement. This work is supported by the National Key Research and Development Program of
China (Nos. 2018YFA0305602 and 2021YFA1401800), the National Natural Science Foundation of China (Nos. 11974404, 11922414, U2032204 and 11888101), the Strategic Priority Research Program (B) of the Chinese Academy of Sciences (Nos. XDB25000000), the Youth Innovation Promotion Association of CAS (No. 2017013), the K. C. Wong Education Foundation (GJTD-2018-01), the Informatization Plan of Chinese Academy of Sciences (CAS-WX2021SF-0102) and Synergetic Extreme Condition User Facility (SECUF). 

\vspace{3mm}

\noindent {\bf Author Contributions}\\

X.J.Z., G.D.L. and C.Y.S. proposed and designed the research. P.B.S. and Y.G.S. contributed to NaAlSi crystal growth. L.J. and X.M.Z. contributed to the band structure calculations and theoretical discussion. C.Y.S., H.T.R., W.P.Z., B.L. carried out the ARPES experiment at NSRL with technical assistance from S.T.C. and Z.S.. C.Y.S., G.D.L. and X.J.Z. analyzed the data. C.Y.S., X.J.Z. and G.D.L. wrote the paper. All authors participated in discussion and comment on the paper.

\newpage

\begin{figure*}[tbp]
	\begin{center}
		\includegraphics[width=1\columnwidth,angle=0]{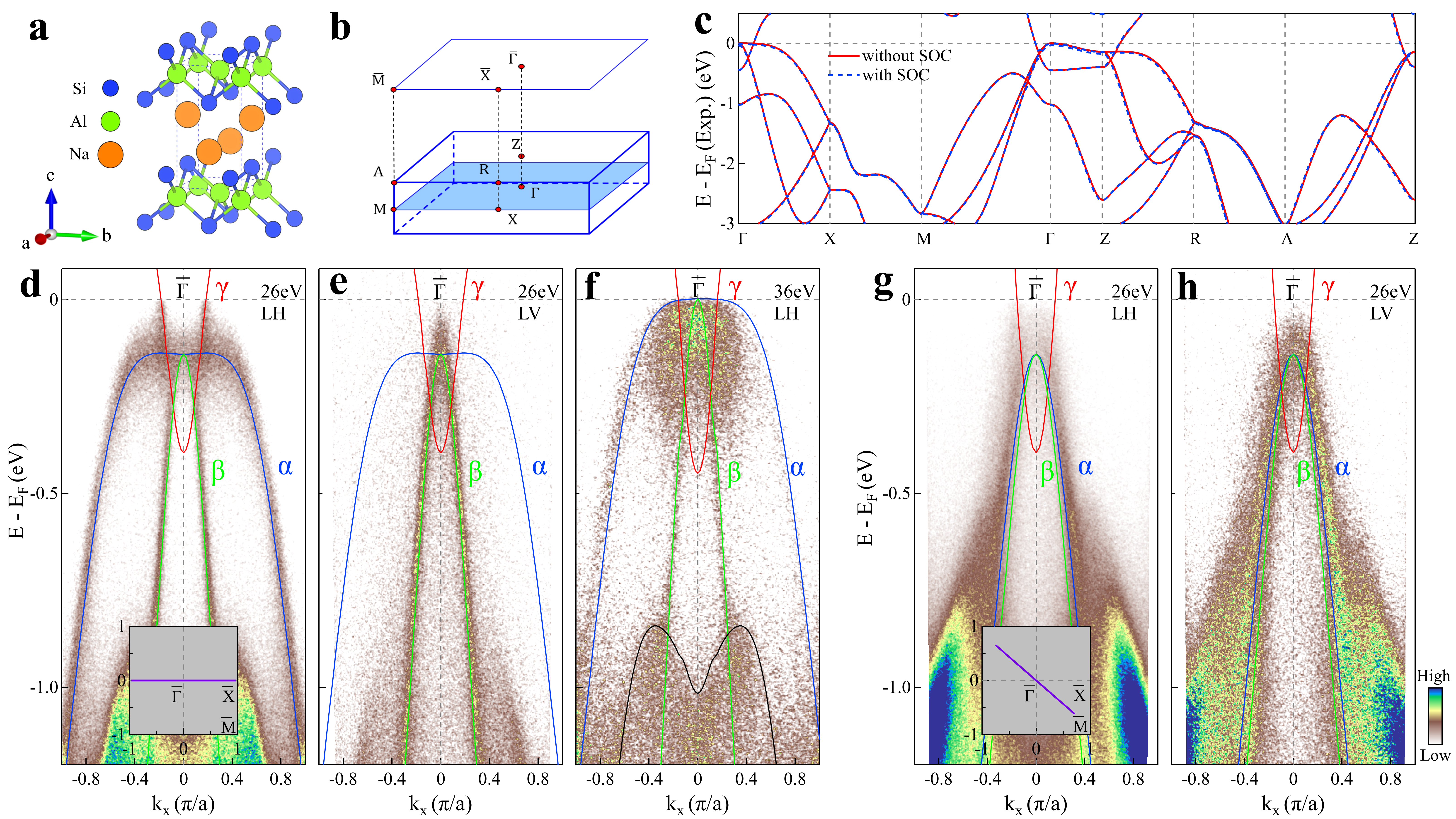}
	\end{center}
	\caption{\textbf{Band structures of NaAlSi and their comparison with band structure calculations.} (a) Crystal structure of NaAlSi. (b) Three-dimensional (3D) Brillouin zone of NaAlSi and the projected two-dimensional Brillouin zone. High symmetry points are marked. (c) Calculated band structures without (red solid lines) and with (blue dashed lines) spin-orbital coupling (SOC). (d) Band structure measured along $\bar\Gamma$-$\bar{X}$ with a photon energy of 26\,eV under the LH polarization geometry. The location of the momentum cut is shown in the inset by the purple line. (e) Same as (d) but measured under the LV polarization geometry. (f) same as (d) but measured with a photon energy of 36\,eV. (g) Band structure measured along $\bar\Gamma$-$\bar{M}$ with a photon energy of 26\,eV under the LH polarization geometry. The location of the momentum cut is shown in the inset by the purple line. (h) Same as (g) but measured under the LV polarization geometry. The observed bands are labeled by $\alpha$, $\beta$ and $\gamma$. The calculated bands are shifted upwards by 0.1\,eV and overlaid on the measured band structures.
	} 
\end{figure*}

\begin{figure*}[tbp]
	\begin{center}
		\includegraphics[width=1\columnwidth,angle=0]{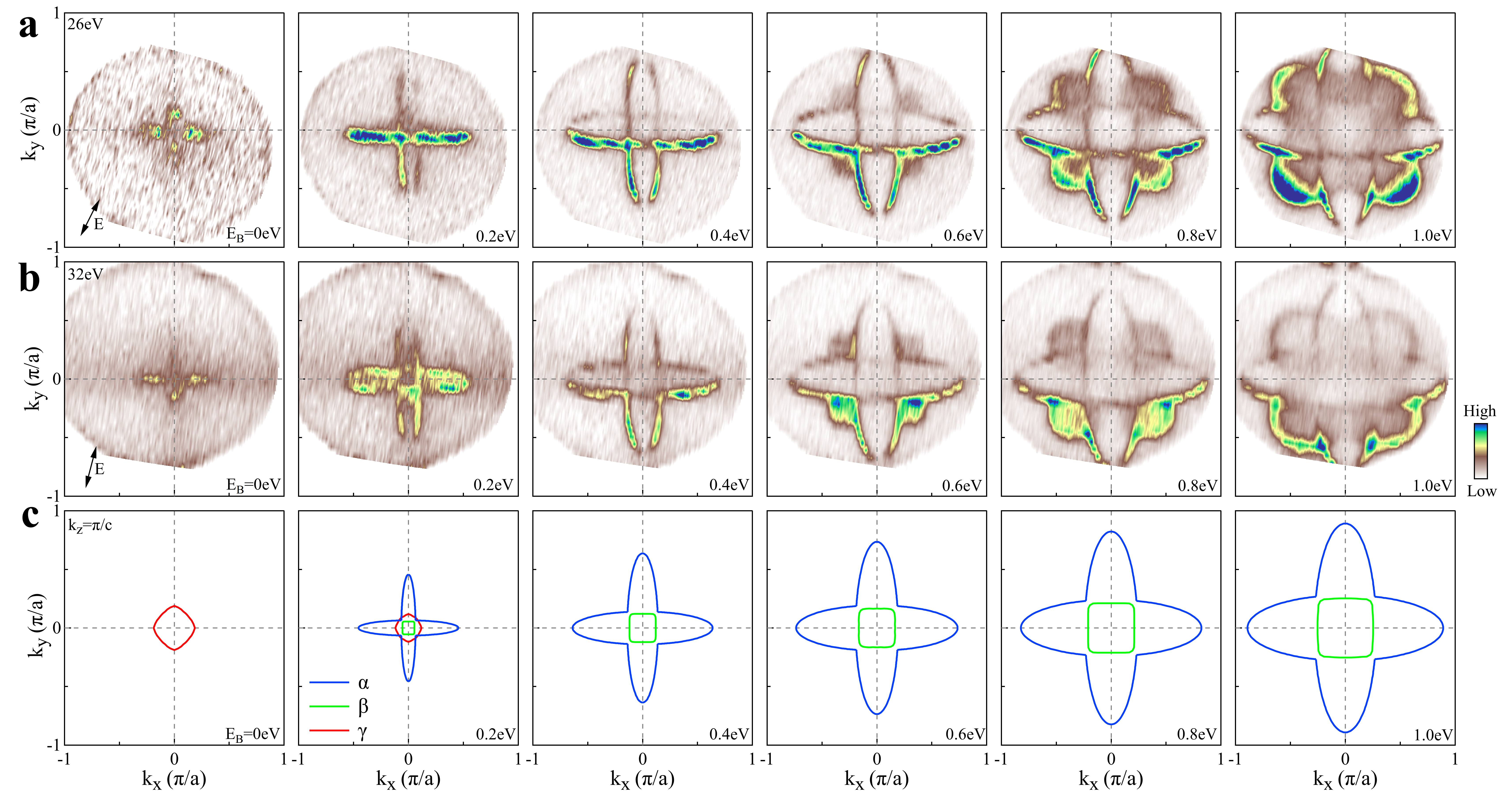}
	\end{center}
	\caption{\textbf{Fermi surface and constant energy contours of NaAlSi and their comparison with band structure calculations.} (a) Fermi surface and constant energy contours of NaAlSi at different binding energies (E$_{B}$) measured with a photon energy of 26\,eV. The major E vector of the incident light is marked at the bottom-left inset of the first panels. (b) Same as (a) but measured with a photon energy of 32\,eV. (c) Calculated Fermi surface and constant energy contours at different binding energies for k$_{z}$ at $\pi$/c.
	} 
\end{figure*}

\begin{figure*}[tbp]
	\begin{center}
		\includegraphics[width=1\columnwidth,angle=0]{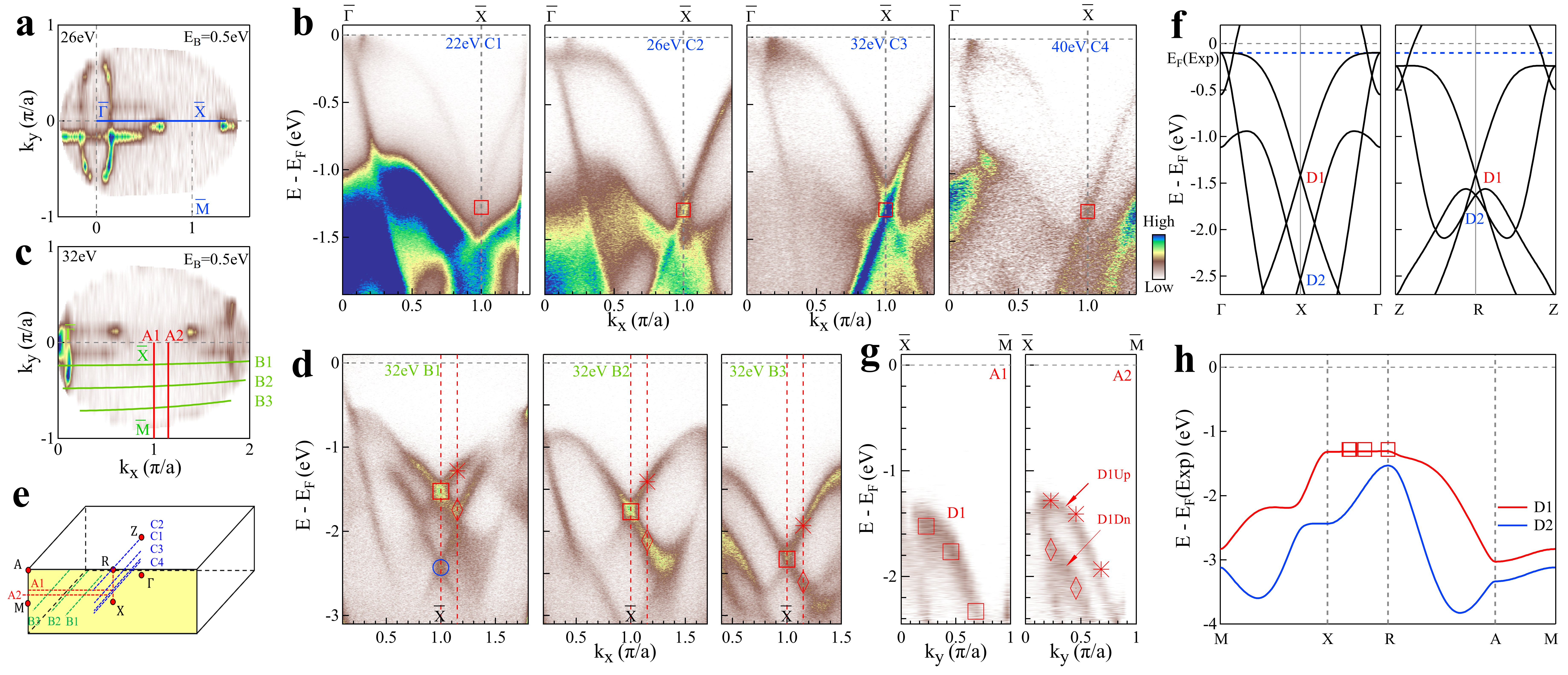}
	\end{center}
	\caption{\textbf{Observation of Dirac nodal surfaces in NaAlSi.} (a) Constant energy contour of NaAlSi at a binding energy of 0.5\,eV measured with a photon energy of 26\,eV. (b) Band structures measured along the $\bar\Gamma$-$\bar{X}$ direction with photon energies of 22\,eV, 26\,eV, 32\,eV and 40\,eV. The location of the momentum cut is shown in (a) by the blue solid line. The positions of the Dirac crossings at $\bar{X}$ are marked by red squares. (c) Constant energy contour at a binding energy of 0.5\,eV measured with a photon energy of 32\,eV. (d) Band structures measured along the $\bar\Gamma$-$\bar{X}$ direction using 32\,eV photon energy. The location of the momentum cuts is shown in (c) by the green solid lines. The positions of the Dirac crossings at $\bar{X}$ are marked by red squares for the ones at lower binding energy and blue circle for that at a higher binding energy. (e) The overall location of the momentum cuts in (b),(d) and (g) in 3D Brillouin zone. The yellow X-R-A-M plane represents the Dirac nodal surfaces. (f) Calculated band structures along $\Gamma$-X-$\Gamma$ (left panel) and Z-R-Z (right panel). The blue dashed lines represent the Fermi level obtained from ARPES measurements. D1 and D2 indicate the Dirac crossings at X and R points. (g) Band structures measured along $\bar{X}$-$\bar{M}$ in the X-R-A-M plane (Cut A1, left panel) and away from the plane (Cut A2, right panel). The locations of the A1 and A2 cuts are marked in (c) and (e) by red lines. In the band structure along A1 cut (left panel), the position of the D1 band obtained from (d) (red squares at k$_{x}$ = 1) is plotted and agrees well with the measured D1 band. In the band structure along A2 cut (right panel), the positions of the two branches of the D1 band obtained from (d) (red asterisks and red diamonds at k$_{x}$ = 1.15) are plotted which agree well with the observed D1Up and D1Dn bands. (h) Calculated band structures without SOC along X-R-A-M lines. The experimental Fermi level (E$_{F}$(Exp)) is used. The locations of the Dirac crossings from (b) are plotted as squares.
	} 
\end{figure*}

\begin{figure*}[tbp]
	\begin{center}
		\includegraphics[width=1\columnwidth,angle=0]{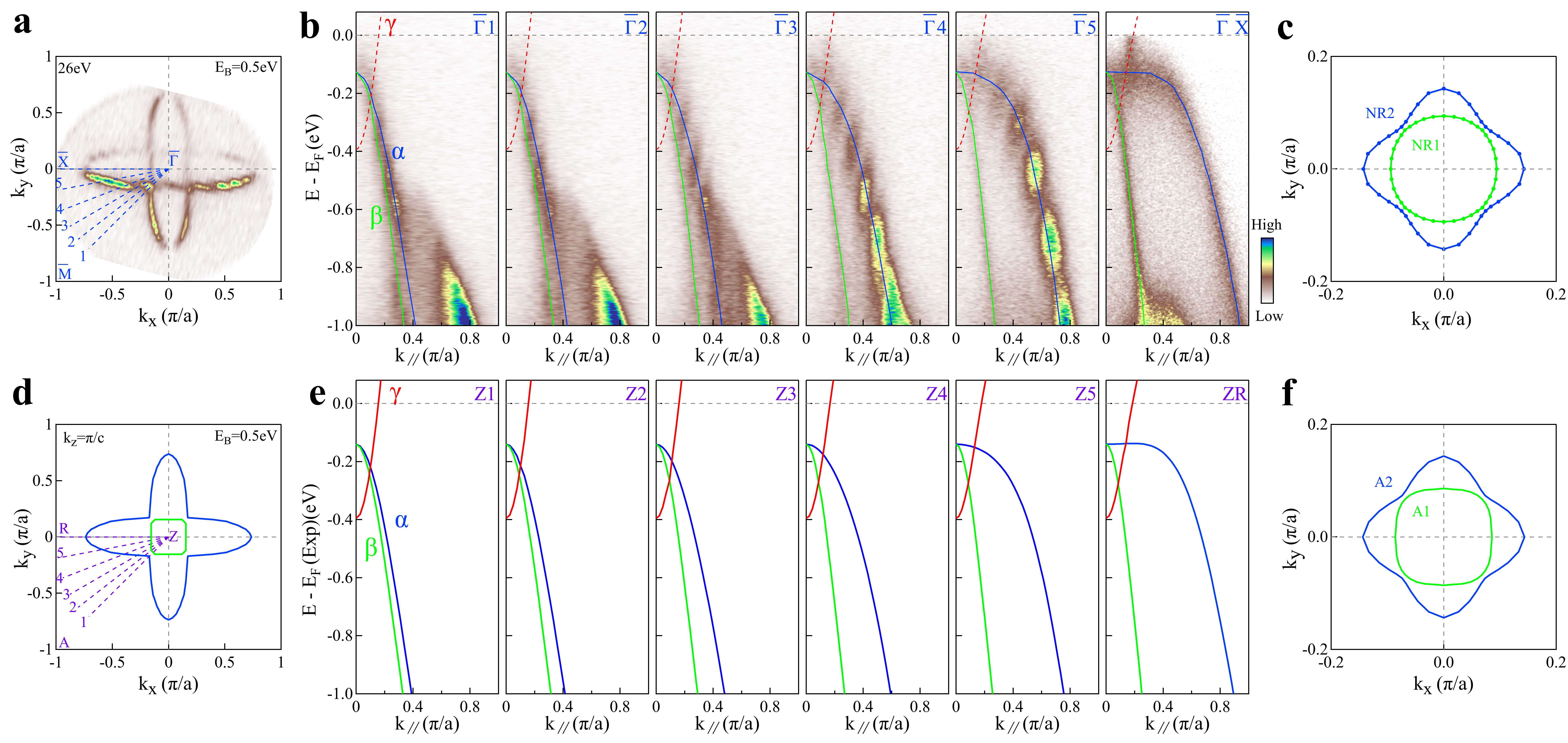}
	\end{center}
	\caption{\textbf{Formation of the Dirac nodal rings in NaAlSi.} (a) Constant energy contour of NaAlSi at a binding energy of 0.5\,eV measured with a photon energy of 26\,eV. (b) Band structures measured along different momentum cuts ranging from $\bar\Gamma$-$\bar{M}$ to $\bar\Gamma$-$\bar{X}$. The location of the momentum cuts is shown in (a) by the blue dashed lines. The blue and green solid lines in (b) are guidelines for the $\alpha$ and $\beta$ bands. (c) The Dirac nodal rings obtained from (b). The inner ring NR1 (green line) is obtained from band crossings between $\beta$ and $\gamma$ bands in (b) while the outer ring NR2 (blue line) is obtained from band crossings between $\alpha$ and $\gamma$ bands in (b). (d) The calculated constant energy contour at E$_{B}$=0.5\,eV at k$_{z}$=$\pi$/c.  (e) Calculated band structures (solid lines) along different momentum cuts ranging from ZA to ZR. The location of the momentum cuts is shown in (d) by the purple dashed lines. (f) The calculated Dirac nodal rings obtained from (e). The inner ring A1 (green line) is obtained from band crossings between $\beta$ and $\gamma$ bands in (e) while the outer ring A2 (blue line) is obtained from band crossings between $\alpha$ and $\gamma$ bands in (e). 
	} 
\end{figure*}

\end{document}